\newcommand{\CLASSINPUTinnersidemargin}{18mm}
\newcommand{\CLASSINPUToutersidemargin}{12mm}
\newcommand{\CLASSINPUTtoptextmargin}{20mm}
\newcommand{\CLASSINPUTbottomtextmargin}{25mm}
\documentclass[10pt,conference,a4paper]{IEEEtran}

\usepackage[utf8]{inputenc}

\usepackage{times}

\usepackage{graphicx}
\usepackage{subfigure}
\DeclareGraphicsExtensions{.png,.eps,.ps,.pdf}

\usepackage{url}
\usepackage[colorlinks=true, linkcolor=black, urlcolor=black, citecolor=black]{hyperref}

\usepackage{amssymb}

\usepackage{booktabs}
\usepackage{amsmath}
\usepackage{graphicx}
\usepackage{multirow}
\usepackage{float}
\usepackage{orcidlink}

\begin{document}

\title{A Novel Protocol Using Captive Portals for FIDO2 Network Authentication}


\DeclareRobustCommand{\IEEEauthorrefmark}[1]{\smash{\textsuperscript{\footnotesize #1}}}

\author{

\IEEEauthorblockN{
\textbf{Martiño Rivera-Dourado}\IEEEauthorrefmark{1}\IEEEauthorrefmark{2}\textsuperscript{*}\orcidlink{0000-0003-4301-9417},
\textbf{Marcos Gestal}\IEEEauthorrefmark{1}\IEEEauthorrefmark{2}\IEEEauthorrefmark{3}\orcidlink{0000-0002-4371-8632},
\textbf{Alejandro Pazos}\IEEEauthorrefmark{1}\IEEEauthorrefmark{2}\IEEEauthorrefmark{3}\IEEEauthorrefmark{4}\orcidlink{0000-0003-2324-238X}, 
\textbf{Jose Vázquez-Naya}\IEEEauthorrefmark{1}\IEEEauthorrefmark{2}\orcidlink{0000-0002-6194-5329}
}\\

\IEEEauthorblockA{\IEEEauthorrefmark{1}Grupo RNASA-IMEDIR, Facultade de Informática, Universidade da Coruña, Campus de Elviña, A Coruña, 15071, Spain}
\IEEEauthorblockA{\IEEEauthorrefmark{2}Centro de Investigación CITIC, Universidade da Coruña, Campus de Elviña, A Coruña, 15071, Spain}
\IEEEauthorblockA{\IEEEauthorrefmark{3}IKERDATA S.L., ZITEK, University of Basque Country UPVEHU, Rectorate Building, Leioa, 48940, Spain}
\IEEEauthorblockA{\IEEEauthorrefmark{4}Biomedical Research Institute of A Coruña (INIBIC), University Hospical Complex (CHUAC), A Coruña, 15006, Spain}%
\IEEEauthorblockA{* \textit{Corresponding author:} {\small \texttt{martino.rivera.dourado@udc.es}}}
}%

\maketitle

\begin{abstract}
FIDO2 authentication is starting to be applied in numerous web authentication services, aiming to replace passwords and their known vulnerabilities. However, this new authentication method has not been integrated yet with network authentication systems. In this paper, we introduce FIDO2CAP: FIDO2 Captive-portal Authentication Protocol. Our proposal describes a novel protocol for captive-portal network authentication using FIDO2 authenticators, as security keys and passkeys. For validating our proposal, we have developed a prototype of FIDO2CAP authentication in a mock scenario. Using this prototype, we performed an usability experiment with 15 real users. This work makes the first systematic approach for adapting network authentication to the new authentication paradigm relying on FIDO2 authentication.\\
\end{abstract}


\section{Introduction}
The society often associates the concept of cybersecurity with passwords. They are omnipresent in information systems all around the world for quite a long time. For a user, they are easy to use and understand and, without a doubt, passwords have become a well-accepted paradigm as an authentication standard.

\subsection{Passwords and their vulnerabilities}
Passwords as an authentication method have become vulnerable to numerous attacks. The most famous is phishing, where an attacker misleads the victim to reveal their credentials. However, phishing is not the only existing threat that the passwords are subject to. Keyloggers are hardware or software tools that attackers use to log the user input into a system. Some hardware keyloggers can be physically placed between the keyboard and the computer to log keystrokes that are directly sent to the attacker. Moreover, it is a reality that information leaks often appear in the news. These leaks can threaten passwords if they contain password hashes used by authentication servers. Although an attacker is not able to directly use password hashes in the authentication portal, the password can be obtained from them with different techniques. Password-cracking techniques are based on automated trial and error methods for finding the original password. 

In conclusion, all phishing, password stealing, and cracking attacks entail a security risk for password-based authentication methods that leave the user unprotected and without control over their credentials. This makes the attacker able to access these systems and, therefore, to steal the information and data they hold.

\subsection{FIDO2: A new standard for web authentication}
Considering that information systems protected by passwords are heavily threatened, some relevant technology companies such as Google, Microsoft, Meta, or Yubico have started to develop a new authentication method. Back in 2014, these companies created the FIDO Alliance \cite{fidoalliance}, which has published two versions of the FIDO Client To Authenticator Protocol (CTAP) standard during the last few years: CTAP1, previously known as Universal Second Factor (U2F), and CTAP2. Since then, some security keys compliant with these standards were developed, such as the Yubikey, the Solokey, or the Google Titan Security Keys.

\begin{figure}[!ht]
    \centering
    \includegraphics[width=\linewidth]{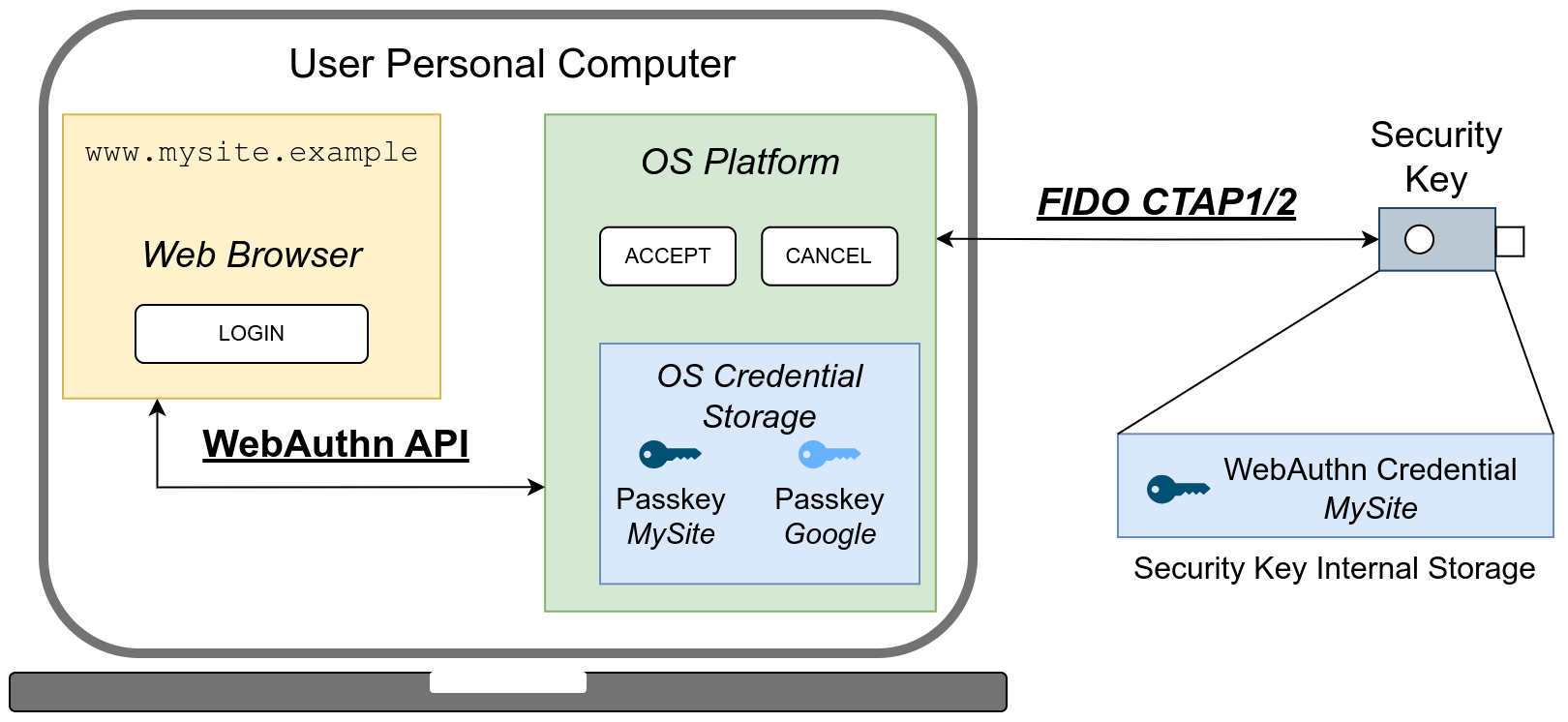}
    \caption{User personal computer compatible with FIDO2: W3C WebAuthn and FIDO CTAP standards. The web application interacts with the OS via the WebAuthn API of the web browser. The personal computer can have a compatible credential storage for WebAuthn credentials, or interact with a security key via FIDO CTAP1/2.}
    \label{img:introduction:webauthn-fido}
\end{figure}

In this context, in 2019, the W3C Consortium developed WebAuthn, a new standard compatible with these hardware security keys~\cite{webauthn-1} and other software-based authenticators. WebAuthn is a new W3C standard that aims to complement or even replace passwords as a web authentication method. Relying on the FIDO standards, WebAuthn defines a browser API that has already been implemented in famous browsers such as Firefox and Chrome~\cite{webauthn-mozilla-table} (see figure~\ref{img:introduction:webauthn-fido}). In addition, it describes a protocol that allows its usage as an authentication method in web application servers. Consequently, some relevant web applications have already added WebAuthn security  keys as an authentication method, to be used together with passwords or even to replace them~\cite{riveradourado-analysis}.

The adoption of WebAuthn has evolved during these few years. ``Passkeys'' are the new commercial name of WebAuthn credentials for passwordless authentication. During the last year, we have seen a great impulse of passkeys, which increased in 50\% from 2019 to 2022 according to a Duo report \cite{duo-report}. For instance, Google~\cite{google-passkeys}, Microsoft~\cite{microsoft-passkeys} and Apple~\cite{apple-passkeys} accounts, some cloud provider portals like Amazon AWS but also in social networks like Facebook and Twitter.

FIDO CTAP standards and the W3C WebAuthn standard form together a new authentication paradigm under the name of \textit{FIDO2 authentication}, also known as the ``passwordless'' paradigm. It opens new possibilities for authentication mechanisms in multiple systems. Although the efforts were directed to web application user authentication, there are other computer systems where it could be integrated. One of them is authentication in computer networks, where access to the network or its resources are controlled.

\subsection{Network authentication}
There are distinct ways to provide access control to a network. Some solutions like EAP with 802.1X can control the connectivity at the link level, but others like captive portal systems perform packet filtering to restrict access to network resources after a client attachment. After authentication, both systems aim to authorise the end device to send and receive traffic on the network.

Wireless home personal networks can use different network authentication protocols for controlling connectivity. Although there has been an evolution in the protocols, from WEP to the last WPA3-Personal, they still rely on long-term passwords to authenticate users.

Meanwhile, many corporate networks use the Extensible Authentication Protocol (EAP) together with the 802.1X standard for authenticating users in wired or wireless networks. These standards are used for controlling the connectivity to the networks at the link level. The most common EAP methods still use passwords as an authentication mechanism, such as EAP-MD5 and LEAAP or some tunnelled methods based on PEAP or EAP-TTLS, like MSCHAPv2 or PAP.

\subsection{Captive portals}
\label{subsec:captive-portals}
Apart from EAP, there exist other alternatives for access control to networks. As mentioned, captive portal systems filter the traffic at the network or link layer to control the access to network resources~\cite{rfc-8952}. Captive portals are websites displayed in end devices after network attachment to an Access Point (AP) before granting access to network resources or the Internet.

There are several ways to implement a captive portal. Most of the solutions nowadays are proprietary implementations delivered in business routers, but other open-source or custom router firmware offer this solution. For this work, we are interested in open-source alternatives for integrating WebAuthn authentication within the captive portal system. The most well-known solutions are pfSense and OpenWRT. They both implement a captive portal system, filtering access to network resources and displaying an captive portal web page to the user. In our work, we will modify this captive portal web page and integrate WebAuthn authentication in this system.

pfSense is one of the most popular open-source firmware available in the market, mainly developed by Netgate. Similar to other proprietary solutions, pfSense offers authentication based on local-configured accounts or an external RADIUS server \cite{pfsense-documentation}. Their firmware is compatible with several routers, including Netgate, but also others like some TP-Link and Protectli models.

On the other hand, OpenWRT is a Linux-based operating system that targets embedded systems \cite{openwrt}. This firmware can be installed in some routers from Lynksis, Asus, TP-Link and Netgear, among others. OpenWRT has many available add-on packages that add features to the core light firmware. One of the available packages, built as a separate open-source project, is OpenNDS. OpenNDS, based on NoDogSplash, is a captive portal software developed in C compatible with this firmware. Among their options, it is worth mentioning it adds the possibility to externalise the web server used in the captive portal. This configuration is known as the Forwarding Authentication Service (FAS) \cite{opennds-fas}.

\subsection{Authenticators: security keys and platforms}
FIDO \textit{security keys} can be used in web authentication thanks to the WebAuthn API, developed by the W3C. These hardware devices are the original idea of having a dedicated hardware known as \textit{authenticator} to generate and store WebAuthn credentials. There are some commercial hardware security keys in the market, like the Yubikey and the Solokey, which can generate WebAuthn credentials. 

Apart from hardware authenticators, there are operating systems that already support creating WebAuthn credentials, which are generated by a software known as a \textit{platform authenticator}. These credentials can be stored in a secure storage at the operating system, or stored in some cloud secure storage.  These are known as ``passkeys''. Passkeys are WebAuthn credentials credentials created in compatible smartphones or laptops~\cite{google-passkeys}, and stored in a cloud provider to be synchronised among different devices of the user.

\subsection{The WebAuthn protocol and types of credentials}
\label{subsec:types-credentials}
Both software-based passkeys and credentials generated in a security key~\footnote{For simplicity, in this work we are also using the term \textit{security key} when we refer to a WebAuthn credential generated and managed by a security key.} are involved in the two main ceremonies of the protocol: registration (or attestation) and authentication (or \textit{assertion}).

During registration, a credential is created and linked with an user account. This involves storing the created public key and the credential identifier in the authentication server. During authentication, the user can verify their identity using the previously registered credential. Specifically, they use the private key to sign a challenge, which is verified by the authentication server.

As said, a FIDO authenticator can generate credentials during the registration ceremony. If the credential is stored into the hardware device memory, then the credential is known as \textit{resident} (or discoverable) credential. If the credential does not use the security key memory, then it is known as \textit{non-resident} (or non-discoverable) credential. For a user, the type of credential is almost transparent. However, from the development point of view, each type of credential involves a different registration and authentication flow. Also, not all security keys and operating systems support resident credentials. In this paper, we developed support for both types of credentials, to increase the compatibility of the system.

\subsection{Related works}
There exist several previous works that have addressed network authentication security improvement before, but there are few of them that have applied FIDO authenticators to network authentication.

In 2018, Chifor \cite{chifor} presented a solution that makes use of the FIDO UAF protocol, a previous version of the current FIDO CTAP protocol, to network authentication. In this work, the protocol is used for authenticating guest users in an enterprise network by employing their personal Android smartphones. For this purpose, they have created a scheme where guests can register themselves to be authenticated in a different Wi-Fi network.

FIDO protocols were also applied to IoT network authentication. Also in 2018, Chifor has proposed a security authorisation scheme for IoT devices, which interact with the user Android smartphone to perform FIDO CTAP1 authentication \cite{chifor-iot}. Then, in 2021, Luo et al. \cite{luo-g2f} have used FIDO CTAP1 security keys to provide user authentication in an IoT Smart Home environment. In their work, a gateway node managing IoT devices authenticates a user when performing management operations. This authentication is performed with FIDO CTAP1 security keys physically connected to the gateway node.

Furthermore, it is worth mentioning the work of Huseynov~\cite{huseynov}, who proposed in 2022 a solution for connecting to a VPN service by authenticating the user with FIDO security keys. Their approach is to create an intermediary web portal that provides a temporal username and password pair, after authenticating with security keys. Their approach is a wrapper solution to the problem, as the authentication of the VPN service is still based on the existing compatible authentication method of the VPN software.

On the other hand, it is interesting how captive portals have been adapted to additional network authentication scenarios. Marquest et al. \cite{eap-sh} designed a custom EAP method named EAP for Secure Hotspots (EAP-SH) that adapts the web-based authentication of Captive Portals to be used in IEEE 802.1X access controls. The work of Marquest et al. can be adapted to use the work of this paper with EAP technology.

Despite the numerous security problems of captive portals~\cite{circumvent-captive-portal}, there have been improvements in this field. The RFC 8910 \cite{rfc-8910} published in 2020 adds a DHCP code to improve the captive portal detection. In wireless networks, the appearance of WPA3, a new Wi-Fi Alliance standard, adds encryption to open Wi-Fi networks with the Opportunistic Wireless Encryption (OWE) protocol \cite{wifi-owe}. This feature mitigates eavesdropping attacks on open wireless networks and, therefore, makes them more secure to use a captive portal on them.

\subsection{Contributions}
In this work, we designed FIDO2CAP: FIDO2 Captive-portal Authentication Protocol. Our proposal is to adapt FIDO2 authentication to the network authentication performed via a captive portal.  Then, we present a prototype of the OpenNDS captive portal using FIDO2CAP. Our system makes the authentication in captive portals resistant to common attacks for which password-based systems are vulnerable. 

For validation, we have installed our prototype in networking hardware. We validated the compatibility with different operating systems and web browsers, and we conducted an usability experiment with 15 users.

Therefore, our contributions can be summarised as follows:

\begin{itemize}
    \item \textbf{A novel protocol that integrates FIDO2 with captive portal authentication: FIDO2CAP}. We provide a design of the FIDO2 Captive-portal Authentication Protocol (FIDO2CAP). We base our architecture design in the captive portal specification RFC 8952~\cite{rfc-8952}, and then we specify the authentication and registration protocols.
    \item \textbf{Prototype of our protocol with OpenNDS}. We have developed, tested and validated with 15 real users a prototype implementation of our protocol. For this, we have used an existing captive portal software known as OpenNDS.
\end{itemize}

\subsection{Paper organisation}
The remaining sections of the paper are organised as follows. Section~\ref{sec:materials-methods} describes the required software and hardware for this work, together with the methodology for all the research, implementation and validation tasks. Then, section~\ref{sec:fido2cap} shows the proposed FIDO2CAP protocol: both the architecture and the message flows. Then, we present a developed prototype of FIDO2CAP in section~\ref{sec:prototype}. We have validated the prototype as explained in the subsection~\ref{subsec:validation-usability}. All the results are presented in section~\ref{sec:results}, which are discussed in section~\ref{sec:discussion}. Finally, we draw some future research lines in section~\ref{sec:future-work}.

\section{Materials and Methods}
\label{sec:materials-methods}
The main objective of this work was to design, develop and validate the integration of FIDO2 authentication with a captive portal, providing a usable system for network authentication. After the FIDO2CAP protocol design, we have developed a prototype. Firstly, we developed an authentication web server and, secondly, we integrated it with the OpenNDS captive portal software. Also, for improving usability, our work involved an study of the exceptions issued by different operating systems, together with a user interface design. Finally, the result is validated with a compatibility test and a usability experiment with real users. This section describes the materials and methods used during all these phases of this work.

\subsection{Required materials}
\label{subsec:materials}
Considering all the involved parties in this development authentication and authorisation system, the environment used during this development tasks is based in Virtual Machines (VMs). Also, we describe the networking hardware used during validation. Additionally, for working with WebAuthn authentication, we also need security keys and a compatible technology stack. Namely:

\begin{itemize}
    \item Virtualisation software. We used \textit{VirtualBox 6.1} for setting up the virtual environment for development.
    \item Three security keys. For the development of the tool and the validation phase, we have used the Yubico Yubikey Security Key, which is compatible with discoverable credentials.
    \item Compatible technology stack. That is, compatible operating systems and browsers. We have used Linux (Manjaro and Ubuntu), macOS Monterey, Windows (10 and 11), Android (v8, v12 and v13) and iOS 16.
    \item Hardware wireless router compatible with OpenWRT. We used ASUS RT-AC1200.
\end{itemize}

\subsection{Work methodology and planning}
This research, development and validation work was planned in six main phases. Namely, (PH1) initial research, for analysing captive portal implementations, WebAuthn technology and configuring the environment; (PH2) protocol design, for grouping the main requirements; (PH3) authentication server development, to develop the web application with WebAuthn authentication, (PH4) system integration, where the developed web application was integrated with the captive portal authorisation; (PH5) usability improvement, where the error handling is designed and user interface is improved; and (PH6) validation, where the result is validated through compatibility tests and usability experiment with real users.

\subsubsection{Initial research}
The initial research includes two main technology stacks: (1) FIDO2: WebAuthn and FIDO authentication; and (2) network authentication with captive portals. Apart from the scientific literature review, this research also included informal tutorials published in technology blogs, open-source projects and market related technology solutions. All these information sources together enrich the research in these technologies, which is continually evolving.

\subsubsection{Development and integration}
When designing the proposed solution, the authors considered two main phases for the development: first, a web authentication server which implements WebAuthn registration and authentication of security keys and passkeys; and second, the integration the WebAuthn authentication with the captive portal authorisation mechanism. These steps correspond to the phases PH3 and PH4 respectively.

\subsubsection{Usability improvement}
For improving usability, the user interface was redesigned after a first test with few users. Also, we noticed that messages displayed originated from different use errors were not user-friendly. Therefore, we identified many possible errors during authentication and described the exceptions issued by web browsers in each operating system, to improve error messages accordingly.

\subsubsection{Validation}
Finally, we have measured system usability by (1) studying the compatibility across the most common operating systems and browsers and; (2) an usability experiment with real users. The compatibility test was tested using the hardware deployment and two already registered security keys, one with discoverable and other with non-discoverable credentials. In relation with the usability experiment, we designed a experiment where the subjects were given the task to connect and authenticate in a Wi-Fi protected with the developed captive portal. Users were observed to identify all use and technology errors, while measuring the time and reactions. After the task, they were asked some questions to measure their satisfaction. Finally, users were asked to repeat the task and to complete a final questionnaire. We have considered usability measures defined by ISO 9241: completeness, efficiency and satisfaction.

\section{FIDO2CAP: FIDO2 Captive-portal Authentication Protocol}
\label{sec:fido2cap}

In this paper, we propose the FIDO2 Captive-portal Authentication Protocol (FIDO2CAP), a captive portal network authentication protocol using FIDO2. This section includes the theoretical design of the protocol. First, the architecture design based on the captive portal architecture defined by RFC 8952~\cite{rfc-8952}. Then, the message flow of the FIDO2CAP protocol.

\subsection{Architecture design}
\label{subsec:fido2cap:architecture}
RFC 8952~\cite{rfc-8952} describes the general architecture of a captive portal in a network. In this work, we adapt and extend this architecture for enabling network authentication using FIDO2 authenticators. This solution can be implemented to provide access control for users in an access-level network, such a Wi-Fi network or a cabled network of end devices.

Figure~\ref{fig:fido2cap-architecture} shows the final architecture diagram, highlighting our proposed changes. In our architecture, we include several elements:

\begin{figure*}[!ht]
    \centering
    \includegraphics[width=1\linewidth]{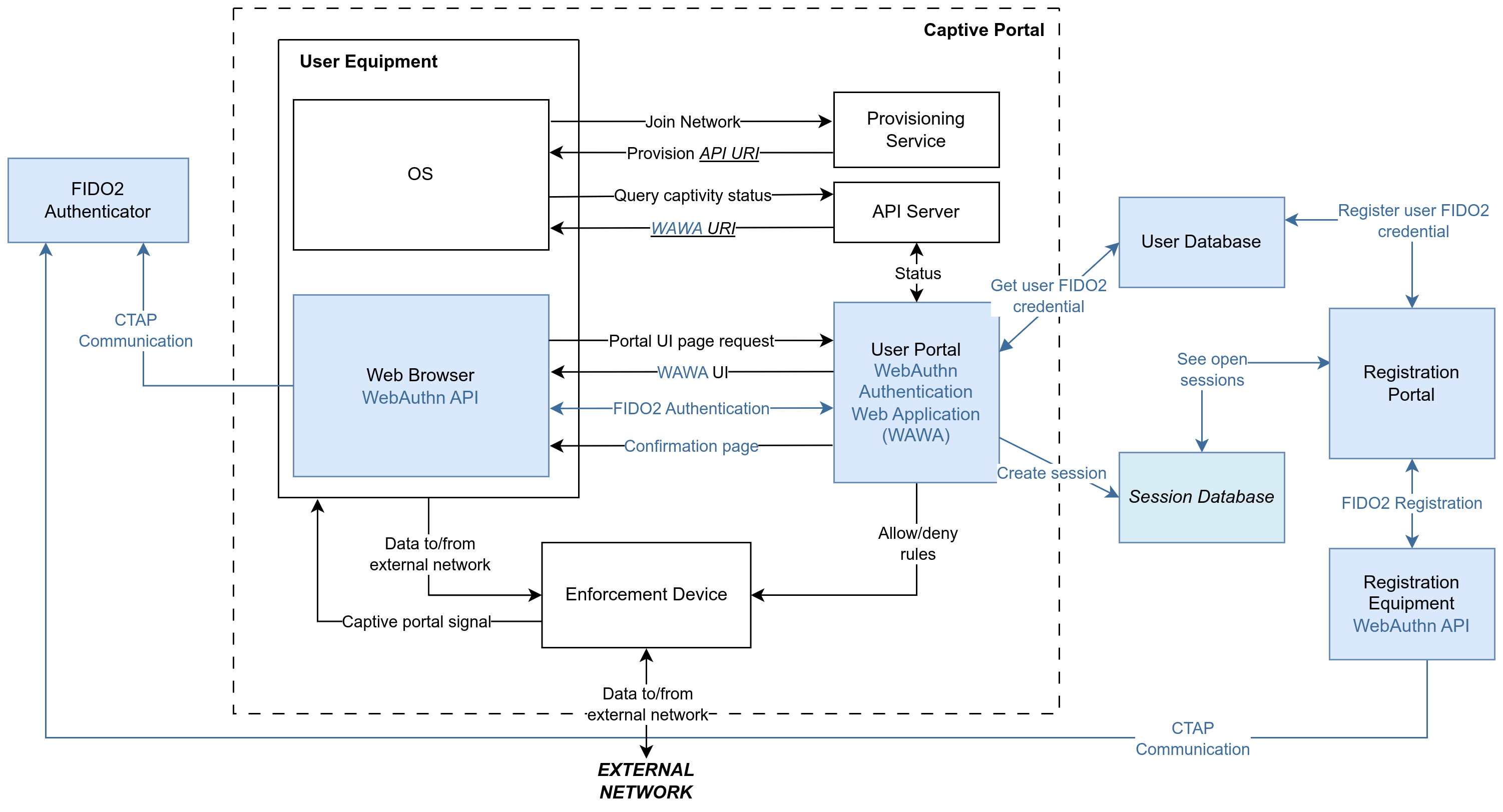}
    \caption{Architecture of the FIDO2 Captive-Portal Authentication (FIDO2CAP) Protocol.}
    \label{fig:fido2cap-architecture}
\end{figure*}

\begin{itemize}
    \item \textbf{FIDO2 Authenticator}. The user FIDO2 authenticator can be a hardware security key, a platform authenticator or a FIDO2 multi-device credential (\textit{passkey}). The user interacts with the FIDO2 Authenticator to prove their identity when accessing a network. A platform authenticator may be included in the user equipment. For describing the general architecture, we consider it a separate device communicating via the FIDO CTAP protocol.
    \item \textbf{Web browser with WebAuthn API}. The user equipment must have a compatible client for performing authentication through the W3C WebAuthn API.
    \item \textbf{WebAuthn Authentication Web Application (WAWA)}. A WebAuthn Relying Party implemented as a web-based application, which works as the user portal displayed by the captive portal system. This includes a web interface displayed at the compatible web browser, and the authentication server.
    \item \textbf{User database}. A database that stores the FIDO2 credentials linked to a user identity.
    \item \textbf{Session database (optional)}. A database of opened sessions, which can allow the user or the administrator to list all open sessions linked with a user identity.
    \item \textbf{Registration portal}. The registration system allows a registrar to register the FIDO2 Authenticator of the user. This portal may also show the active sessions linked with a user and an authenticator. Alternatively, the registration portal may provide a self-registration flow for a user to register themselves. For instance, this self-registration can be controlled by a temporal access token, shared as a QR code.
    \item \textbf{Registration equipment}. The registration equipment and the user equipment may be different, depending on the registration flow. For example, if the registration process is done by an administrator, or there is a setup equipment for self-registration.
\end{itemize}

The proposed system architecture works similarly to the RFC 8952~\cite{rfc-8952} for network authentication. After joining the network, the captive portal intercepts the traffic and provisions the WAWA URI. Then, the Operating System of the User Equipment displays the WAWA User Interface to the user. The WebAuthn Authentication Web Application now starts the FIDO2 authentication process, involving the user and the FIDO2 Authenticator via the WebAuthn API and the FIDO CTAP protocol. If the authentication is successful, the Enforcement Device is instructed to allow the access to the External Network for the User Equipment. This authorises the user to access the protected network.

\subsection{Protocol overview}
\label{subsec:fido2cap:protocol}
The FIDO2CAP (FIDO2 Captive-portal Authentication Protocol) describes two main ceremonies, following the FIDO2 standards. Firstly, we describe the user authentication protocol, based on the proposed architecture. Then, we propose a FIDO2 registration flow via the Registration Portal.

\subsubsection{Registration scenarios: administrator or self-registration}
\label{subsec:fido2cap:self-registration}
We consider two main actors in the protocol: an user and a registrar. Also, we consider a different equipment for authentication (User Equipment) and for registration (Registration Equipment). This provides a flexible range of environements. For instance, the system may be enabled for user self-registration. The user and the registrar may be the same person. In this case, the Registration Portal by a temporal access token for user self-registration, which may be shared as a QR code to the user. This would allow the user to use the User Equipment for registration, which enables the registration of user FIDO2 Platform Authenticators. For example, a user may scan a QR code provided by an administrator, and use their Android smartphone to register a passkey in the system.

\subsubsection{FIDO2 Credentials: discoverable and non-discoverable}
\label{subsec:fido2cap:credentials}
As mentioned in section~\ref{subsec:types-credentials}, FIDO2 credentials can be discoverable (resident) or non-discoverable (non-resident). This changes how the authentication protocol works. With non-discoverable credentials, the user must be identified in the first place, while discoverable credentials do not need this step. When the user is identified (for instance, using a username), the list of allowed credentials is sent to the FIDO2 authenticator. As non-discoverable credentials are not resident (stored) at the authenticator, they need to be provided in the allowed credentials list during the authentication protocol. Our definition of FIDO2CAP protocol is compatible with discoverable and non-discoverable credentials, as it includes the identification of the user. If discoverable credentials are used, then this step during authentication can be omitted.

\subsection{Authentication protocol}
\begin{figure*}
    \centering
    \includegraphics[width=1\linewidth]{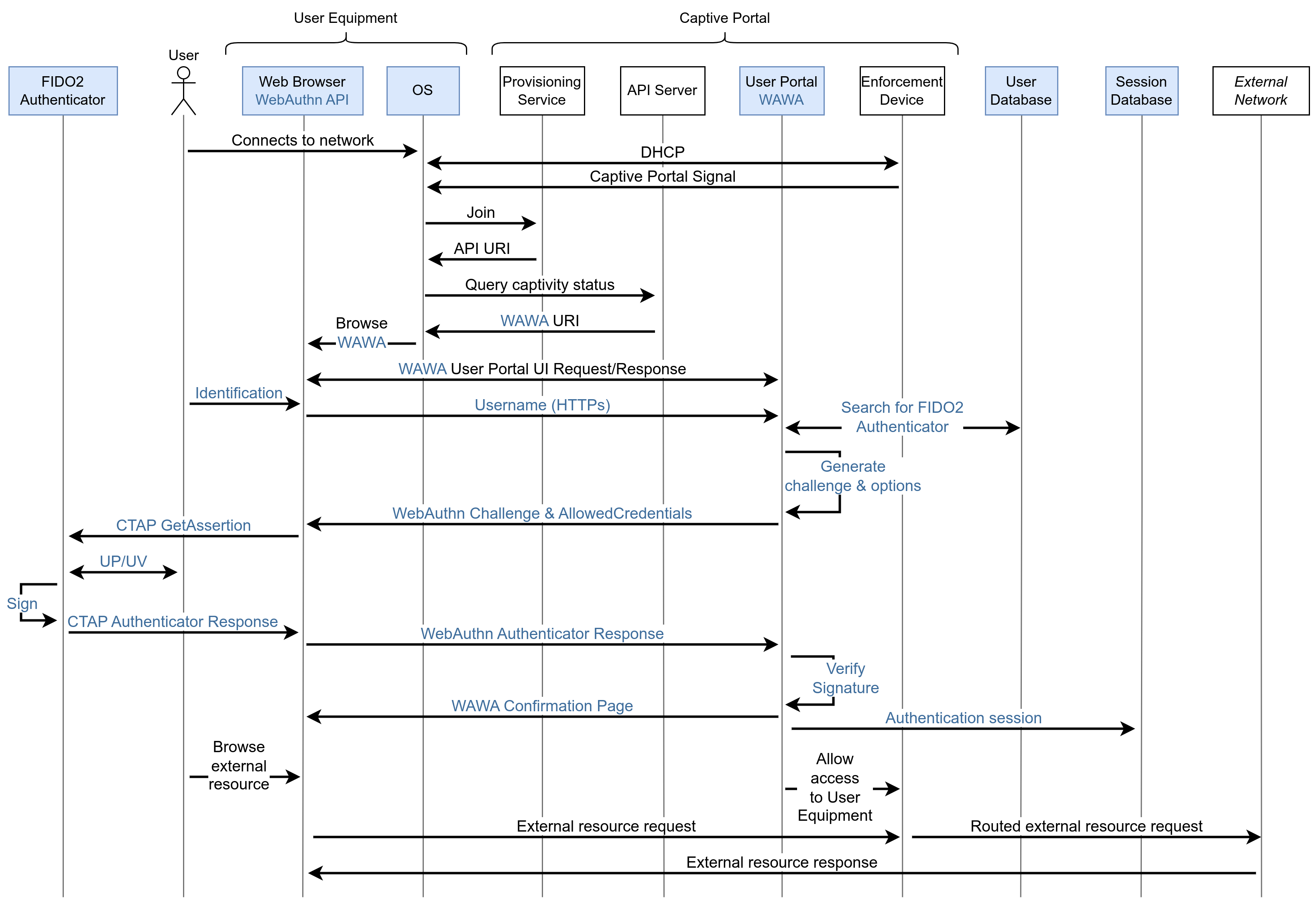}
    \caption{Authentication message flow in FIDO2CAP. The User Equipment is connected to the Captive Portal, getting redirected to the WAWA server URI. The user performs FIDO2 authentication at the WAWA User Portal UI. After verifying the authentication, the Enforcement Device allows the access to User Equipment, ending captivity.}
    \label{fig:fido2cap-authentication-flow}
\end{figure*}

Figure~\ref{fig:fido2cap-authentication-flow} shows the FIDO2CAP authentication protocol message flow. The parts highlighted in blue are modified or added to the RFC ~\cite{rfc-8952}. FIDO2CAP protocol starts with the captivity signaling from the router. When the user equipment requests the captivity status, the captive portal API Server replies with the WAWA URI. This opens a web browser in the user equipment, visiting the WAWA user interface. Here, the user sends the username for identification (this is the optional step with discoverable credentials, see section~\ref{subsec:fido2cap:credentials}). 

The username allows the WAWA application to search for registered FIDO2 authenticators linked with the user in the User Database. With this information, the WAWA server generates the Allowed Credentials list, and the WebAuthn challenge, which is sent to the Web Browser. In the user end, the browser initiates the WebAuthn API, which inteacts with the authenticator via the OS CTAP GetCredentials call. After the user interacts with the authenticator via User Presence (UP) and/or User Verification (UV), the FIDO2 Authenticator signs the challenge and sends its response, which is forwarded to the WAWA server. Finally, the response is verified, and the WAWA server notifies the Enforcement Device, allowing the User Equipment to access the external resources, ending the captivity.

\subsection{Registration protocol}
\begin{figure*}
    \centering
    \includegraphics[width=0.65\linewidth]{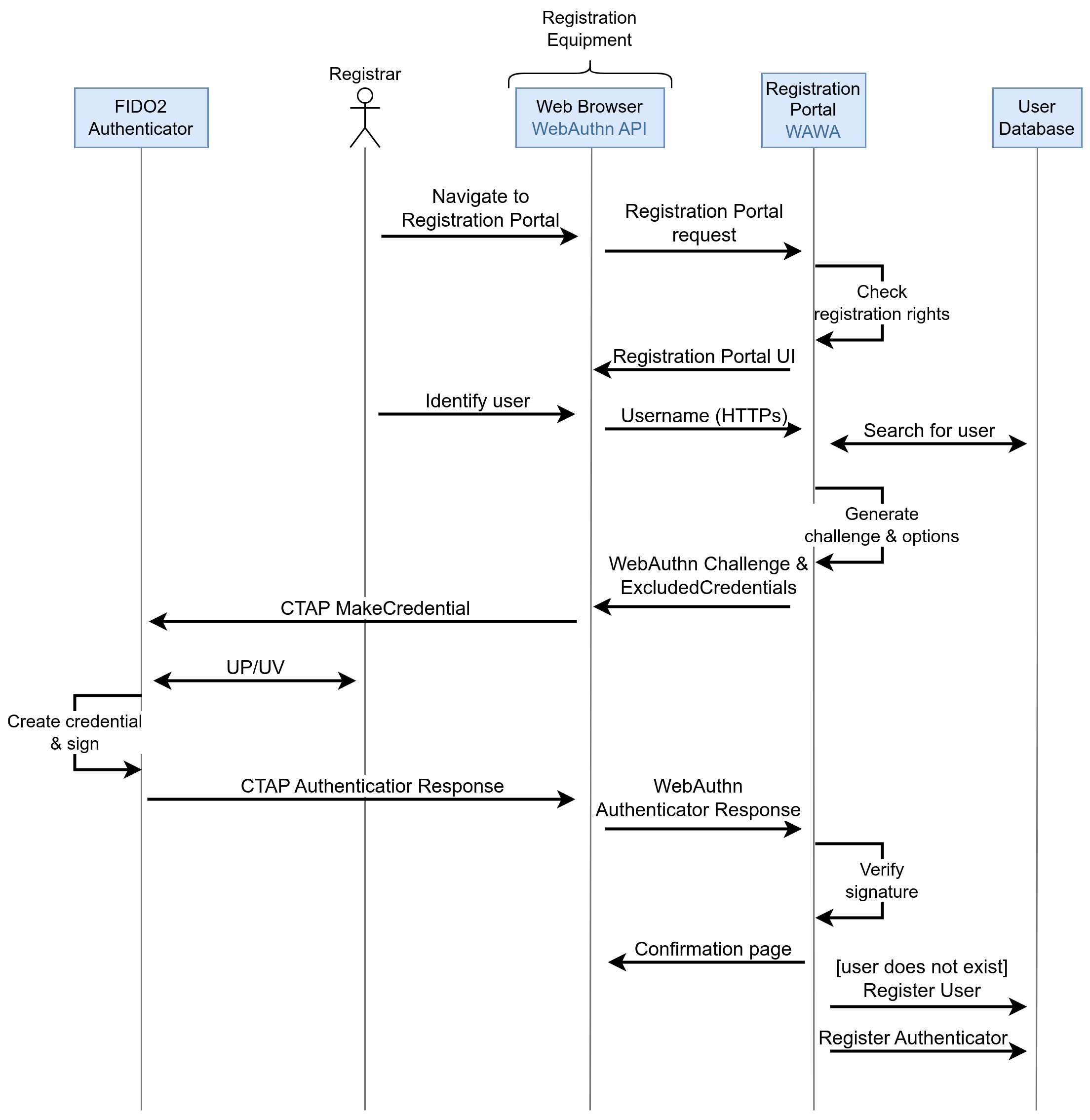}
    \caption{Registration message flow in FIDO2CAP. The Registrar has the required permissions to access the Registration Portal UI. After specifying the username, FIDO2 registration starts. Finally, the FIDO2 credentials (referred as Authenticator) are registered linked with the user account.}
    \label{fig:FIDO2CAP-registration-flow}
\end{figure*}

FIDO2CAP registration protocol can be instantiated with different setups. Figure~\ref{fig:FIDO2CAP-registration-flow} shows the general registration process, which involves the WAWA Registration Portal, a different user interface of the WAWA server. The Registrar, which may be the user, sends a request to the Registration Portal. After verifying the Registrar has the required registration rights (see section~\ref{subsec:fido2cap:self-registration}), the Registrar sends the user username. Then, the username is used to search for registered FIDO2 Authenticators in the User Database. These are be included in the Excluded Credentials list, together with the registration challenge generated by the WAWA server. This triggers the WebAuthn API that interacts with the user FIDO2 Authenticator via the CTAP MakeCredential command. Once requested, the Registrar authorises the registration in the authenticator, which generates the response. This response is forwarded to the Registration Portal, who verifies it. Then, if the user is already registered, the credential is added to the allowed FIDO2 Authenticators. If the user is not present in the database, it is registered with the new credential. 

\section{Prototype development and validation}
\label{sec:prototype}
As a Proof of Concept (PoC), we have developed and validated a prototype of the proposed FIDO2CAP protocol. In this section, we present the developed prototype: a captive portal authentication system using FIDO2CAP. First, we explain the WAWA server implementation (section~\ref{subsec:prototype:wawa}), and then the integration with the OpenNDS captive portal software (section~\ref{subsec:prototype:integration}).

Our prototype follows the FIDO2CAP architecture (see figure~\ref{fig:fido2cap-architecture}). We implemented the WAWA server, using a single database as the User Database and Session Database. This web application includes both a User Portal and a Registration Portal, which implement the WebAuthn functionality. Finally, we have integrated the our WAWA implementation with OpenNDS. OpenNDS implements a captive portal: the Enforcement Device, the Provisioning Service and the API Server.

\subsection{WebAuthn Authentication Web Application (WAWA)}
\label{subsec:prototype:wawa}
In this section it is described the WebAuthn Authentication Web Application (WAWA) that allows (1) registration of WebAuthn credentials by the administrator and (2) the authentication with a WebAuthn credential by the user. The following sections include relevant details on the development, which was based in \textit{NodeJS} and the \textit{SimpleWebAuthn} library~\cite{simple-webauthn}.

\subsubsection{WebAuthn standard and credentials}
As explained in section~\ref{subsec:types-credentials}, the WebAuthn standard considers registration and authentication of credentials. There are two types of WebAuthn credentials related to authenticators (like security keys): discoverable and non-discoverable. Discoverable credentials are stored physically within the authenticator. On the contrary, non-discoverable credentials are stored encrypted in the server. In this prototype, both discoverable and non-discoverable credentials have been considered. For this purpose, we have to consider two different authentication flows.

\subsubsection{Authentication with discoverable credentials}
With discoverable credentials, the user does not need to identify themselves with a \texttt{username} during the process. The \texttt{userHandle} included in the authenticator response is used as the \texttt{userId} to find the registered device for verifying the authenticator response. Hence, using discoverable WebAuthn credentials implies that the user is identified \textit{after} the authentication of the credential.

\subsubsection{Authentication with non-discoverable credentials}
On the contrary, with non-discoverable credentials, the authentication response from the authenticator will not contain a \texttt{userHandle} that identifies the user. Therefore, the \texttt{username} should be specified at the authentication form. When the server generates the WebAuthn authentication options, a list of the registered credentials for the corresponding user is provided. This list contains credential identifiers that are actually encrypted private keys that only the registered authenticator can decrypt to correctly authenticate the user.

\subsubsection{Administration interface}
The registration of credentials in the web application is restricted to administration. Using a RBAC, whether the users are authorised to register new devices or see restricted information depends on their role. For this purpose, we created a separate administrator interface. The interface is restricted to users with the \textit{admin} (or registrar) role, which  authorises a user to: (1) check registered users, their active sessions and registered credentials; (2) register a WebAuthn credential; and (3) assign administration roles to users.

\subsubsection{Session management}
\label{subsec:sessions}
Web sessions represent an authenticated user and links their HTTP requests with that identity, so they can be authorised. In this web server, \textit{ExpressJS} uses cookies to maintain that session and expires them within a configured time, specified in an environment variable.

For administrators to enumerate active sessions for the users, the sessions have also been stored in the web server database. That way, from the administrator interface, the administrator can list the active user sessions per registered user. That is also useful when integrating with the captive portal to authorise authenticated users. Each time the user is authenticated in the server, a new session is stored linked with the user identifier. With this approach, a user can have multiple concurrent sessions, which are represented in the administration interface. When integrating the server with other systems like the captive portal, users are able to authenticate more than once without requiring them to logout first from other sessions.

\subsection{Captive portal integration with OpenNDS}
\label{subsec:prototype:integration}
Once we developed a first working version of the WAWA, it was integrated with a captive portal software: \textit{OpenNDS}. This software implements the Provisioning Service, the API Server and the Enforcement Device. First, we explain why we have chosen OpenNDS for the prototype development. Secondly, we explain the integration of the WAWA with the API Server and the Enforcement Device of OpenNDS.

During this section, the captive portal is commonly referenced with the chosen solution name: \textit{OpenNDS}. On the other hand, our WAWA server is commonly referenced as \textit{FAS server} (Forwarding Authentication Service), nomenclature used in the OpenNDS configuration for the User Portal server.

\subsubsection{Selection of a captive portal solution: OpenNDS}
Section~\ref{subsec:captive-portals} shows the existing captive portal technology nowadays. The integration work was successfully achieved with OpenNDS installed on a Open-WRT firmware. However, the first tests were performed with pfSense.

Both pfSense \cite{pfsense-documentation} and OpenNDS \cite{opennds-fas} are captive portal solutions that are suitable for this integration for being open-source. Therefore, they could be modified to integrate the WebAuthn authentication server to achieve a final WebAuthn captive portal. However, the design of both solutions is different. While pfSense is a complete router firmware that implements a captive portal, OpenWRT \cite{openwrt} router firmware relies on the OpenNDS independent package to implement the captive portal functionality. This makes OpenNDS a more self-contained solution whose project code is easier to approach.

A captive portal includes several key elements: an Enforcement Device, an API Server, the User Portal and an the Provisioning Service. One specific advantage of OpenNDS that determines the final selection of the solution, is that it allows externalising User Portal of the captive portal. This way, our WAWA server can be modified to be integrated with OpenNDS.

OpenNDS feature for externalising the User Portal is called \textit{Forwarding Authentication Service (FAS)} \cite{opennds-fas}, which forwards the redirected requests to a captive portal to the selected external web server. Finally, OpenNDS enforcement device should be integrated with the external web server to identify the authenticated client and authorise its access to the network. Therefore, during this section, the developed WebAuthn Authentication Web Application (WAWA) server was modified to implement an external \textit{FAS server} compatible with OpenNDS.

\subsubsection{OpenNDS Authmon: API Server and Enforcement Device}
OpenNDS has a module named \textit{Authmon} that implements an API Server and an Enforcement Device. As implemented in OpenNDS, the \textit{Authmon} module sends periodic requests to the FAS server. In this prototype, the FAS server is our WAWA server, which we have made compatible by extending the WAWA API, accepting the requests from \textit{Authmon}.

\begin{figure*}[!ht]
    \centering
    \includegraphics[width=0.6\textwidth]{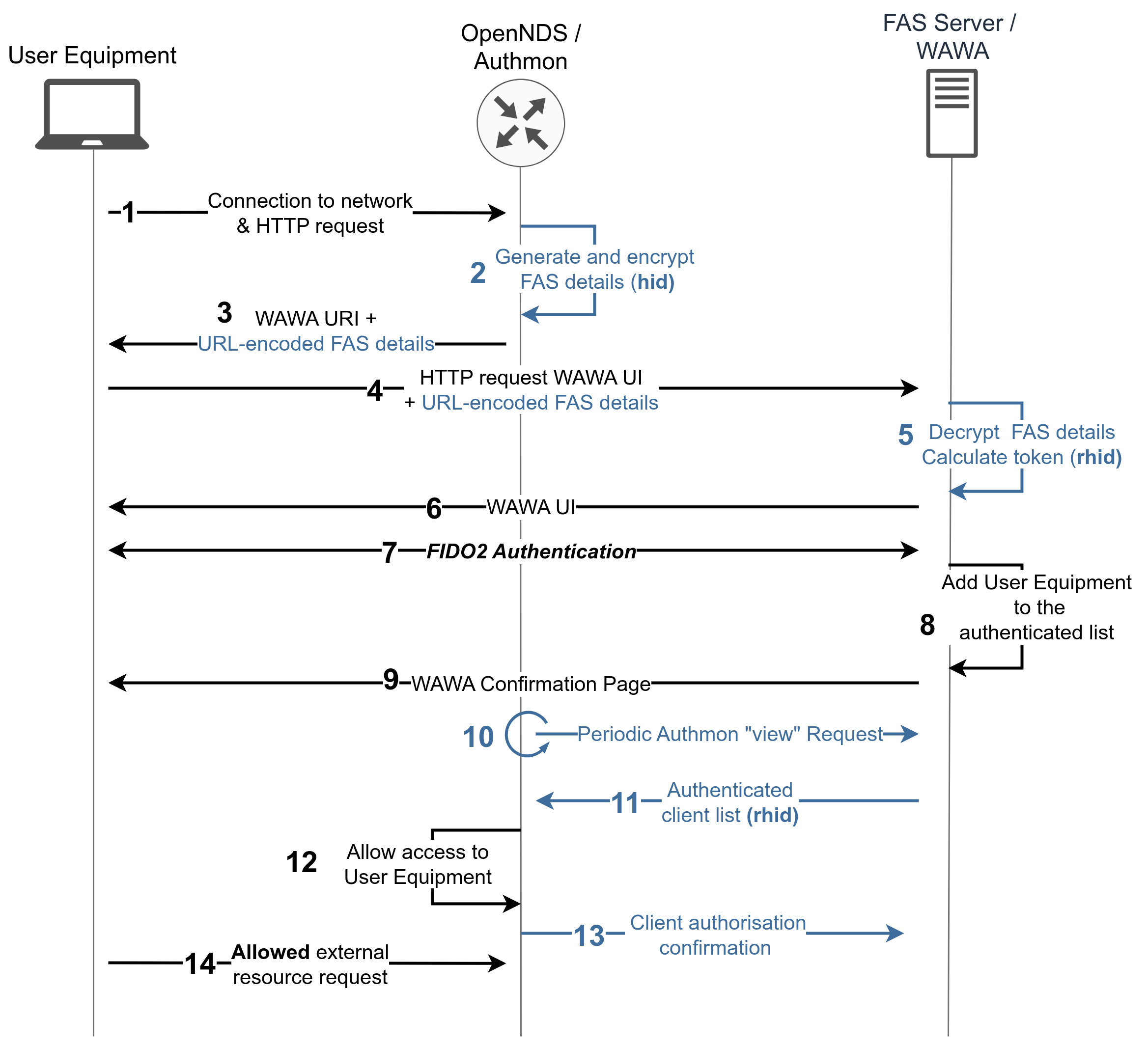}
    \caption{Communication between OpenNDS Authmon module and the developed WAWA server (FAS server). The User Equipment, gets redirected to the WAWA UI hosted at the FAS server and, once authenticated, it is authorised by OpenNDS. In blue, the particular operation of OpenNDS we considered for the integration of the developed WAWA server.}
    \label{img:integration:flow-diagram}
\end{figure*}

The \textit{Authmon} module of OpenNDS sends periodic requests to the FAS server to poll for new authenticated clients. Figure~\ref{img:integration:flow-diagram} shows a flow diagram that represents the final integration of the \textit{Authmon} operation to authorise clients with the developed FAS server. It is worth mentioning that this integration was possible after the reverse engineering process due to the lack of detailed documentation.

Following figure~\ref{img:integration:flow-diagram}: (1-3) the captive portal run by OpenNDS redirects unauthorised clients to the WAWA server URL, together with a hashed id (\texttt{hid}), encoded and encrypted as explained in section~\ref{subsec:cryptography-notes}; (4-5) when handling User Equipment requests, the WAWA server must decrypt these details and calculate the authorisation token \texttt{rhid}, as explained in section~\ref{subsec:cryptography-notes}; (6-9) when the FIDO2 authentication is successful, the WAWA server marks the client as authenticated; (11) all authorisation codes (the \texttt{rhid}) are requested periodically to the WAWA (FAS) server, who returns a list of authorisation tokens is sent to the OpenNDS router; (12-14) using this token, OpenNDS can authorise the corresponding client for their access to network resources, notifying WAWA about this. 

\subsubsection{New API endpoints for Authmon}
Thanks to the reverse engineering, the WAWA REST interface was adapted to handle the \textit{Authmon} periodic requests and the new User Equipment (client) requests. According to the reverse engineering process, OpenNDS \textit{Authmon} module can issue three HTTP POST requests to the authentication WAWA (FAS) server. These are distinguished by the \texttt{auth\_get} parameter of the HTTP request body.

\begin{itemize}
    \item \textbf{''clear''}: All authenticated clients should be clear from the list. That is, the list is reset. This is used by OpenNDS when booting.
    \item \textbf{''list''}: The list should be sent and cleared. This type of request is not frequent and is kept for backwards compatibility.
    \item \textbf{''view''}: The most often request depends on its payload:
    \begin{itemize}
        \item \textbf{* or none}: The complete list of authenticated clients is required by Authmon. The corresponding authorisation tokens (or \texttt{rhid}) should be sent in a list, according to the compatible format: \texttt{* <rhid>}.
        \item \textbf{* \textless~rhid~\textgreater}: Authmon is confirming the authorisation of a client.
    \end{itemize}
\end{itemize}

To implement the list of authenticated clients, the session management feature developed in the WAWA server can be used. As explained in section~\ref{subsec:sessions}, the server incorporates a separate database collection for storing authenticated user sessions. Then, once OpenNDS confirms the client authorisation, it is marked as authorised. Finally, the session expiration time is synchronised with OpenNDS expiration time.

\subsubsection{Cryptographic notes on FAS in OpenNDS}
\label{subsec:cryptography-notes}
For the implementation of the integration there are two cryptographic particular notes that should be considered: (1) the initial encrypted details and (2) the authorisation token of OpenNDS captive portal.

Steps 3 to 6 of figure~\ref{img:integration:flow-diagram} show how the client traffic is redirected to the captive portal at the FAS Server. OpenNDS adds some parameters necessary for the later client authorisation,  encoded in base64 and encrypted with AES-256-CBC. Therefore, the WebAuthn authentication server installed at the FAS Server should decode and decrypt these parameters. Specifically, the AES block cipher in \textit{Cipher-Block-Chaining} (CBC) mode used by OpenNDS has a 256-bit block length. In the WebAuthn authentication server in NodeJS, the required key length is of 32 bytes, which is shared with OpenNDS.

On the other hand, step 5 of figure~\ref{img:integration:flow-diagram} show the calculation of the authenticated hash performed by OpenNDS. This authenticated hash has to be used by the FAS server to send an authorisation token that allows a client to be authorised by OpenNDS. The hash used by OpenNDS is SHA256. However, in order to authenticate the hash, OpenNDS developers have chosen to include the symmetric key at the end of the payload to hash. In this case, the FAS server should return a re-hashed version of the \texttt{hid} parameter when the client is authenticated. Letting $k$ be the 32 byte shared key, the operation is shown in eq.\ref{eq:rhid}.

\begin{equation}
\label{eq:rhid}
rhid = sha256\, (\, hid \: || \: k \,)
\end{equation}

\subsection{Test deployment and improvements}
\label{subsec:deployment}

This section depicts the environment setup of the developed prototype: the WAWA server integrated with OpenNDS. The environment presented here is the one used in this work to perform development and validation of the prototype. Our setup provisions a  Wi-Fi connection via a WPA3 open network. This network is protected with our prototype captive portal solution. It only provides Internet connectivity after successful FIDO2 user authentication.

\subsubsection{Environment configuration of a mock use case: Wi-Fi connectivity in a hotel}
\begin{figure}[!ht]
    \centering
    \includegraphics[width=1\linewidth]{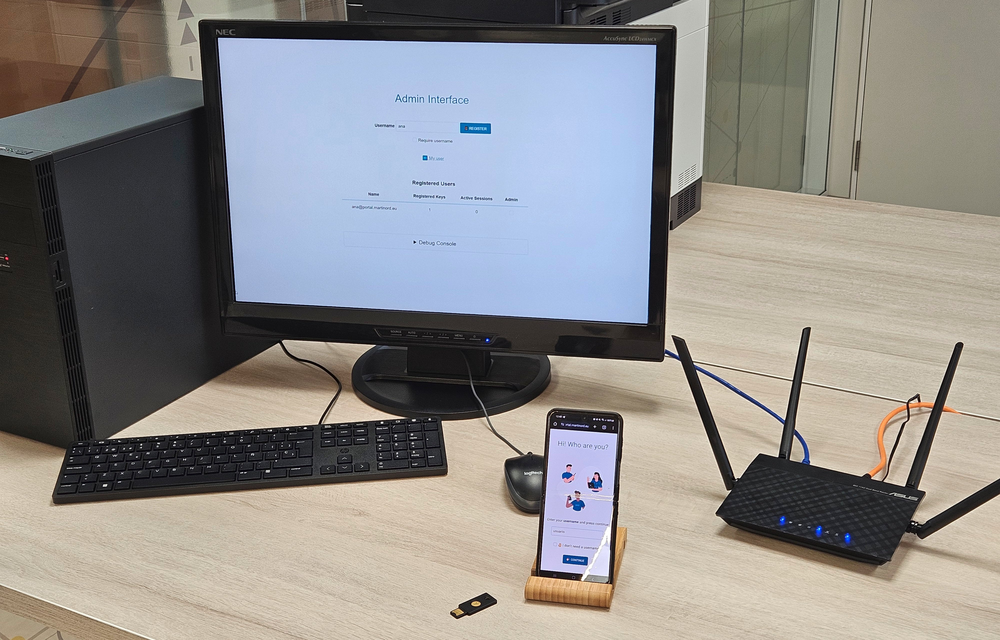}
    \caption{Physical deployment of the example environment. From left to right: the Registration Equipment (Windows 10), the FIDO2 Authenticator (Yubikey Security Key), the User Equipment (Android 13) and the ASUS Router (OpenWRT). The WAWA server is publicly hosted at our servers (running the prototype \cite{fido2cap-server-github}).}
    \label{fig:environment:deployment}
\end{figure}

Our environment setup can be summarised in four equipments, mocking an use case of a hotel providing connectivity to the hotel network for their hosts. The FIDO2 security keys are registered beforehand, and they are provided to the clients. As shown in figure~\ref{fig:environment:diagram} and~\ref{fig:environment:deployment}, the setup consists in: 

\begin{figure}[!ht]
    \centering
    \includegraphics[width=1\linewidth]{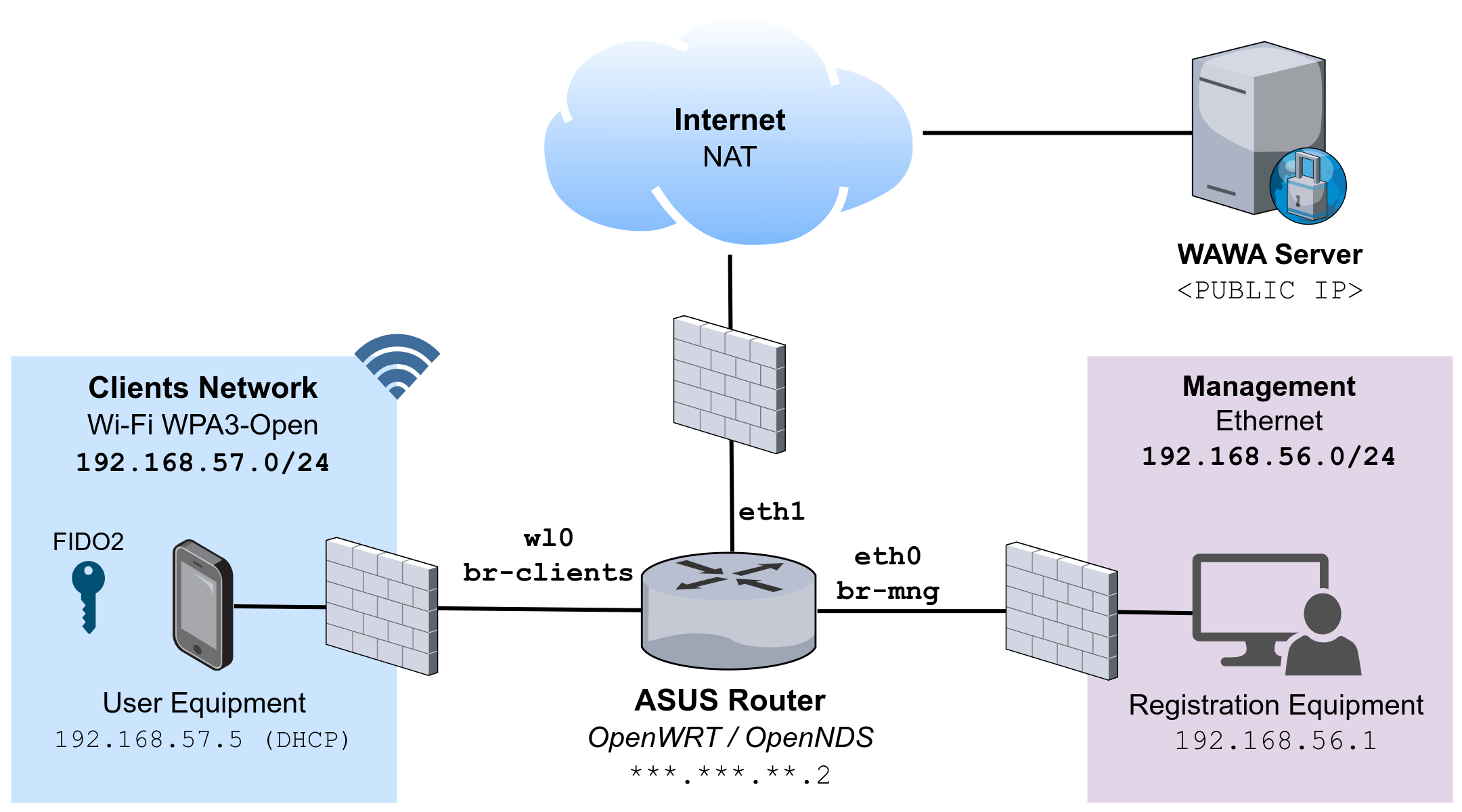}
    \caption{Example environment diagram: Wi-Fi connectivity in a hotel using the developed prototype. The network is divided in three zones: (1) a clients wireless network; (2) a management network; (3) Internet through NAT. The WAWA Server is placed in a public server for simplicity. Clients have restricted access to Internet until successful FIDO2CAP authentication.}
    \label{fig:environment:diagram}
\end{figure}

\begin{enumerate}
    \item The \textbf{User Equipment} and the \textbf{FIDO2 security keys}. They are the personal devices of the hotel clients, used to connect to the Wi-Fi for Internet connectivity.
    \item The \textbf{Registration Equipment} for the administrator. Using this device, the hotel personnel register security keys for their clients.
    \item The developed \textbf{WAWA server}. Running in a physical server, the WAWA application is configured properly to work with the OpenNDS instance.
    \item The \textbf{OpenWRT router running OpenNDS}. For validating our solution with real hardware, we have used the ASUS RT-AC1200 router, which is compatible with OpenWRT. OpenNDS should be properly configured to work with the WAWA. 
\end{enumerate}

The environment was configured by installing the OpenWRT firmware in the ASUS router, using the firmware restoration tool. We have configured a WAN interface as a DHCP client for provide Internet connection to the router through another network. Also, the client network was configured using Wi-Fi with WPA3 open network.

Moreover, as shown in figure~\ref{fig:environment:diagram}, the example topology divides the corporate networking into three networks attached to the OpenWRT router, namely: (1) Internet, (2) management and (3) clients network. For simplicity, our WAWA server is placed in Internet, although it could be in a DMZ within the corporate network. Following our mock example of a hotel network, the clients are restricted to access Internet and some hotel services until successful connection to the Wi-Fi network. This provides the proper segmentation of the network, and access control to clients via FIDO2CAP authentication.

Finally, as mentioned before, configuring the captive portal in the router relies in OpenNDS configuration with the WAWA server. As said, in our example environment the WAWA server is deployed in a public domain with a specific unique IP address. An unique IP address is necessary so OpenNDS can configure the Enforcement Device (or firewall) properly, to allow the User Equipment initial request for the WAWA UI. Also, we apply compatible configuration of the server and the router, so the router can serve as a gateway of the authentication server. Although we include a more detailed version in Github~\cite{fido2cap-server-github}, here are the main configuration steps:

\begin{enumerate}
    \item Install OpenWRT in the ASUS compatible router~\cite{openwrt}.
    \item Install OpenNDS in OpenWRT
    \begin{enumerate}
        \item Configure the clients network.
    \end{enumerate}
    \begin{enumerate}
        \item Download OpenNDS and edit the configuration file (see step 4).
    \end{enumerate}
    \item Install WAWA in a server.
    \begin{enumerate}
        \item Download the WAWA server code from Github~\cite{fido2cap-server-github}.
        \item Configure the WAWA server integration with OpenNDS (see step 4).
    \end{enumerate}
    \item Verify the configuration compatibility: OpenNDS and WAWA:
    \begin{enumerate}
        \item The FAS remote IP:port = WAWA IP:port.
        \item The FAS FQDN as the domain name of WAWA.
        \item The same FAS key random secret of 32 bytes.
        \item The same session timeout.
    \end{enumerate}
\end{enumerate}

\subsubsection{Improvements in usability and compatibility of the prototype}
During the deployment and initial testing of the prototype, we found some usability and compatibility issues. Here we explain some of them, and in the following section we include the final system validation and usability experiment.

During FIDO2 authentication, there are several errors that can occur, so the user should be notified. However, we found that not all errors are properly notified to the user through their web browser. This decreases the usability of the authentication flow UI. In fact, the browsers rise different browser exceptions upon the same error. For instance, Firefox and Chrome in Android differ when the user disconnects a FIDO2 security key during authentication. At the time of writing, Linux rises "Unknown Error" exception, while the Chrome browser rises "Not Readable Error".

For this reason, we are conducting a study of all these differences, and finding some guidance, which will be published in the near future. In the prototype proposed in this paper, we have addressed some of these issues. Some of the errors display a non-intrusive message, like a "Try again" button, without identifying the error and allowing the user to repeat the action. Other errors that occur when the operation was cancelled, include this information to the user, allowing them to do another try.

\subsection{Prototype validation and usability}
\label{subsec:prototype:validation}

A captive portal using FIDO2 authentication is a novel approach. For this reason, we have measured both software compatibility with real devices and system usability with real users of the developed prototype. This section describes both studies.

\subsubsection{Captive portal with WebAuthn compatibility test}
\label{subsec:validation-compatibility}
Our captive portal prototype has been tested in several desktop and mobile operating systems. Taking into account the most used operating systems nowadays \cite{statista-1}  \cite{statista-2}, we have conducted some specific compatibility tests for validating the solution. We have tested Windows, macOS and Linux desktop OSs; and Android, iOS and Samsung One UI mobile OSs. Namely, the following tests were designed:

\begin{itemize}
    \item NET-1 (CPD): The captive portal is detected by the OS and a browser is launched.
    \item NET-2 (INTERNET): There is internet connection after successful captive portal authentication.
    \item WA-1 (WEBAUTHN): The browser launched by the OS is compatible with WebAuthn, the API exists.
    \item WA-2 (NDC): The browser was able to successfully use non-discoverable credentials (non resident).
    \item WA-3 (DC): The browser was able to successfully use discoverable credentials (resident).
    \item WB-1 (REDIRECT): In case of not compatibility of the default browser for captive portal detection, the user is correctly redirected to other browser.
    \item WB-2 (EXCEPTION): Exception issued when the default browser has no practical WebAuthn compatibility.
\end{itemize}

Table~\ref{table:compatibility} shows the test results. The tests were performed with the ASUS router running the captive portal. We registered two FIDO2 security keys, one with discoverable credentials and other with non-discoverable credentials. As shown in the table, there are different operating systems that detect the captive portal and launch a specific mini-browser, that usually is not compatible with WebAuthn. As analysed in \cite{wang_mini_captive_browsers}, these captive portal mini browsers have limited capabilities, and even security issues. These browsers are used in smartphone OSs and Linux-based desktop OSs, which affect to our system compatibility.

\subsubsection{Usability of the captive portal with security keys in mobile devices}
\label{subsec:validation-usability}
As part of the validation of the solution, we designed and conducted an usability experiment with 15 subjects. For this experiment, subjects used their own smartphone devices, which they are used to, as clients of the Wi-Fi network running our captive portal. This follows the mock environment described in section~\ref{subsec:deployment}.

Users were asked to connect to the Wi-Fi network and authenticate with the security key and an username, after a brief explanation. The security key and username were already registered in the system by an administrator. After the first try, the subjects were asked about their satisfaction and completeness of the process. Then, they were asked to perform the task two more times and finally complete a small questionnaire.

For measuring the usability of the solution, we have considered three different measures based on ISO 9241 \cite{iso-9241}: effectiveness, completeness and efficiency:

\begin{itemize}
    \item Effectiveness ``is the accuracy and completeness with which users achieve specified goals", that is, that complete the connection to the Wi-Fi network with the security keys. Also, we measure the error rate while performing the task, by observing the difficulties or technical issues that the user encounters.
    \item Efficiency ``is the resources used in relation to the results achieved", that we measure as time-based efficiency and overall relative efficiency.
    \item Satisfaction ``is the extent to which the user's physical, cognitive and emotional responses that result from use of a system, product or service meet user's needs and expectations". We have measured satisfaction with a custom questionnaire after performing the task.
\end{itemize}

\subsubsection{Usability experiment subjects and devices}
In this experiment, 15 users whose age varies from 18 to 64 have participated in the experiment, including men an women. All of them have a daily use of technology, and half of them have high or expert knowledge about computer science. Also, although the 93\% use second-factor authentication, only the 20\% have tried security keys before. For this reason, one of the hypothesis of the experiment is that the first contact with the proposed system will delay the completion of the task.

The users have used their own smartphone devices, 13 of them Android and 2 iOS. We should highlight that iOS devices are not compatible with the system, as the captive portal is opened in a non-compatible web browser. Most of Android devices had the Android 12 or 13 version, which allow the user to open an alternative compatible browser for authenticating in the network. As all users have daily technology use, they should be able to manipulate their own devices correctly, which helps them to perform tasks reliably.

\section{Results}
\label{sec:results}
This work resulted in a novel network authentication protocol based on FIDO2 authentication: FIDO2CAP. The protocol includes a new architecture based on RFC 8952~\cite{rfc-8952}, and a message flow design for authentication and registration. Our proposed architecture adds a WebAuthn Authentication Web Application (WAWA), a FIDO2 authenticator and other elements to a captive portal. Also, we present a prototype implementation of FIDO2CAP in an environment mocking a hotel Wi-Fi network. Finally, we conducted compatibility tests and a usability experiment with 15 real users.

In this section, we summarise the specific results, namely: (1) the FIDO2CAP protocol; (2) a prototype of the FIDO2CAP in a mock scenario; (3) results of the compatibility tests and the usability experiment.

\subsection{FIDO2CAP: a novel protocol for captive-portal FIDO2 authentication}
\label{subsec:results:fido2cap}
In section~\ref{sec:fido2cap} we define FIDO2CAP: FIDO2 Captive-portal Authentication Protocol. This is the main result of this paper. The proposed novel protocol is based on captive portals defined by RFC 8952~\cite{rfc-8952}, and adapted to support FIDO2 user authentication. The main advantage of this approach is that any captive portal compatible with the RFC could be modified with the architecture of FIDO2CAP to support this new authentication protocol.

FIDO2CAP architecture is shown in figure~\ref{fig:fido2cap-architecture}, where we have included some elements, like the User Database and the WAWA. On the other hand, we have considered the two FIDO2 ceremonies: authentication and registration. The message flows are presented in figure~\ref{fig:fido2cap-authentication-flow} for authentication, and figure~\ref{fig:FIDO2CAP-registration-flow} for registration. In both flows we take into account discoverable and non-discoverable credentials, which ensures compatibility with different FIDO2 authenticators.

\subsection{Functional prototype of FIDO2CAP}
\label{subsec:results:prototype}
We have developed a prototype of FIDO2CAP in the scenario described in section~\ref{subsec:deployment}. This scenario describes a hotel Wi-Fi network protected by a FIDO2CAP captive portal system, which controls access to Internet and other network resources to the hotel clients.

The developed prototype prototype has been published as open-source in Github \cite{fido2cap-server-github}, available to download and deploy in a compatible OpenWRT router. It provides documentation of the configuration options and some instructions for its deployment. In this section, we describe the specific results.

\subsubsection{WebAuthn Authentication Web Application (WAWA)}
WAWA is a web application that  allows a user to authenticate using FIDO2 authenticators compatible with the WebAuthn standard, using a web browser. Registration of users, and their corresponding security keys, are done by a registrar, which is a privileged user implemented using RBAC roles.

\begin{figure}[!ht]
    \centering
    \includegraphics[width=0.7\linewidth]{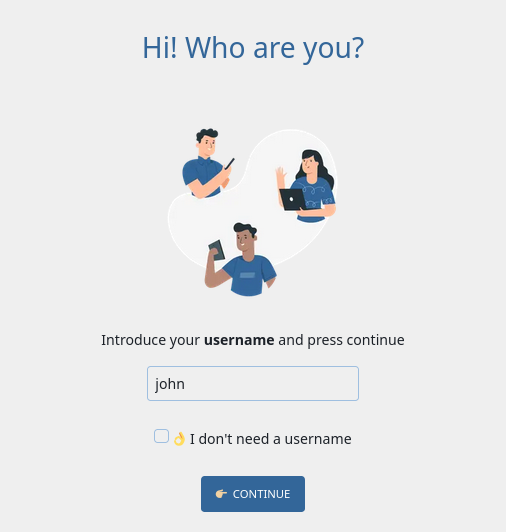}
    \caption{Developed Captive Portal UI. The WebAuthn login form uses discoverable credentials when the `I don't need a username" option is selected.}
    \label{img:results:captive-portal-ui-non-discoverable}
\end{figure}

The developed server is compatible with the last W3C Recommendation of the WebAuthn standard L3 \cite{webauthn-3}, implementing both discoverable and non-discoverable WebAuthn credentials. This feature makes all types of FIDO2 security keys or software ``passkeys" compatible with the server registration and authentication procedures. In the developed system, modern ``passkeys" or discoverable credentials are supported, so the user does not need to introduce an username in the authentication form. If the user has software or hardware not compatible with discoverable credentials, they can use the system with non-discoverable credentials by introducing their username (see figure~\ref{img:results:captive-portal-ui-non-discoverable}).

\begin{figure}[!ht]
    \centering
    \includegraphics[width=0.9\linewidth]{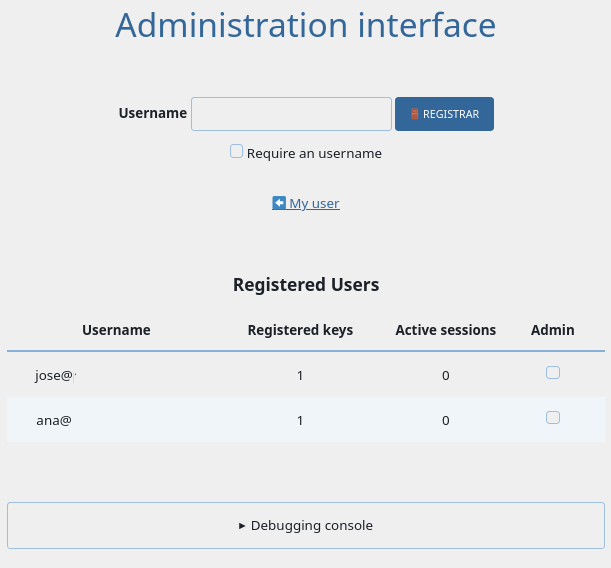}
    \caption{The administration interface of the developed authentication server. It allows registration of security keys. The table of registered users lists the registered device of each user and the active sessions in real time. Admin column is used to configure admin privileges to users.}
    \label{img:results:captive-portal-ui-administrator}
\end{figure}

Additionally, an administrator user in the server can list authenticated users in real time (see figure~\ref{img:results:captive-portal-ui-administrator}). The session database allows users to open more than one authenticated web session, which are listed in the administration panel. Although the user can logout at any time during the session, the expiration time will force the end of the web session automatically.

Finally, an administrator can register security keys. These can be associated with an existing user by specifying the username, or can be registered and associated to a new user. Therefore, a user can have multiple registered security keys. This feature allows users to have a backup security key, which can be used in case of device loss to securely gain access to the network. 

\subsubsection{Integration of WAWA in the OpenNDS captive portal}
\label{subsec:results:integration}
The FIDO2CAP prototype includes the integration of the WAWA server with the OpenNDS captive portal network authentication system. This integration adapted the WAWA server to communicate with the OpenNDS Enforcement device and API Server, as defined by the FIDO2CAP protocol. The integration allows the authorisation of the User Equipment traffic after the FIDO2 authentication with the WAWA server. The following are some specific results achieved during the development of the FIDO2CAP prototype.

Firstly, with this integration, our prototype benefits from the compatibility of OpenNDS. For instance, the integrated solution triggers the Captive Portal Detection (CPD) technology embedded in the most used operating systems, showing the captive portal page once connected to the network.

Other relevant results is that the same WAWA server can be configured to work with different OpenWRT routers (or access points) to control access in different networks. This way, the resulting prototype provides a central point of authentication compatible with a multi-gateway scenario. The same user can be authenticated in different networks, creating different sessions that can be viewed from the administration panel, while all gateway requests are managed independently.

Finally, thanks to the integration with OpenNDS, the solution is compatible with a real deployment scenario, that we used to validate the system (see section~\ref{subsec:results:validation}). OpenNDS can be installed on OpenWRT router firmware, which can form part of a final production environment easy to deliver with multiple hardware routers. These devices can run the OpenNDS captive portal software, configured to be used with the integrated developed WAWA server. This way when a client connects to the network via OpenWRT, it gets redirected to the WebAuthn authentication server, who manages the request accordingly. After the client authentication, the authentication server waits for the periodic requests and confirms the client authorisation, redirecting them to the original HTTP request.

\subsection{FIDO2CAP compatibility and usability}
\label{subsec:results:validation}
With the described FIDO2CAP prototype deployment, we have tested different operating systems and browsers. Finally, we have conducted an usability experiment with real users. These results provide evidence of the feasibility of the protocol and its usability. In this section we include the compatibility results, and then the usability experiment results.

\subsubsection{Compatibility with operating systems and web browsers}
\begin{figure}[!ht]
    \centering
    \includegraphics[width=0.55\linewidth]{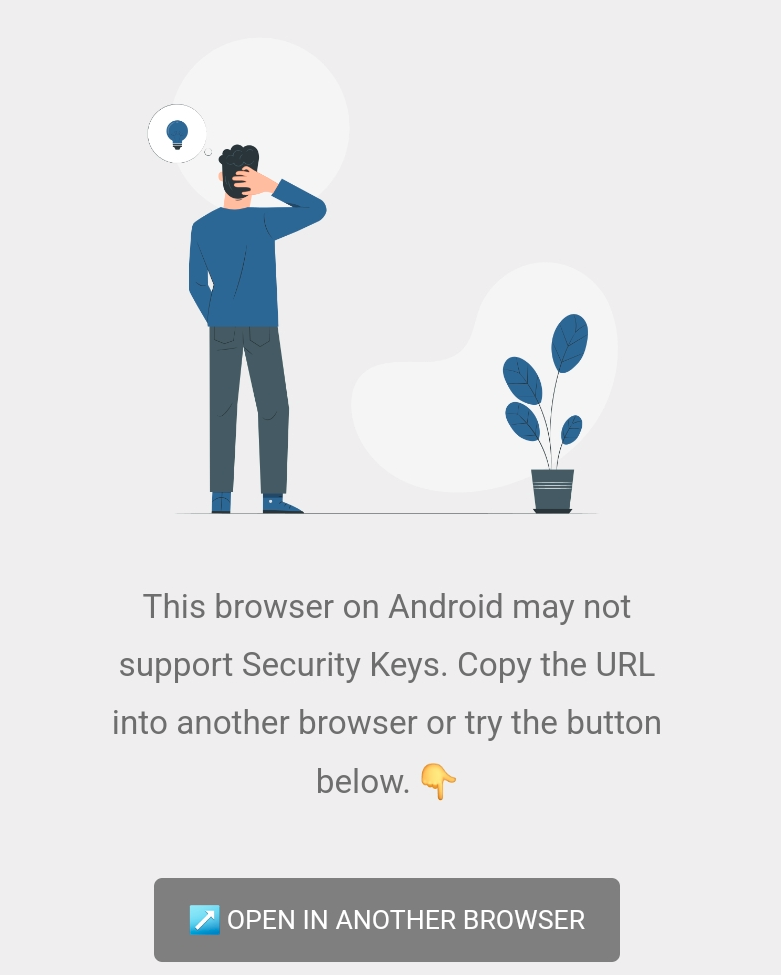}
    \caption{Prototype UI warning the user that the OS or the browser are not compatible. The button opens a new browser. The target browser depends on the specific OS running the request. This screenshot was done in an Android 13 phone running the default mini-browser for captive portals, and the button should open the Google Chrome app with the same URL.}
    \label{fig:results:browser-not-compatible}
\end{figure}

Applying FIDO2 authentication to a captive portal is a novel approach. For this reason, there are some captive portal specific browsers that operating systems launch that do not support the WebAuthn API. We have tested our prototype solution in the most common operating systems and browsers. Table~\ref{table:compatibility} shows the compatibility validation results. As it can be seen, the system is fully compatible with operating systems using the user-defined web browser, which usually supports WebAuthn. There are other operating systems that use a custom web browser with limited functionality to open captive portal web pages, known as mini-browsers~\cite{wang_mini_captive_browsers}, which mostly are not compatible with WebAuthn. When there is no compatibility in a captive portal mini-browser, the developed web interface shows a redirection button to the user, which opens a compatible web browser in Android and iOS (see figure~\ref{fig:results:browser-not-compatible}).

\begin{table*}[!ht]
\resizebox{\textwidth}{!}{%
\def\arraystretch{1.3}%
\begin{tabular}{ccccccccccc}
\hline
\multicolumn{1}{|c|}{}                                     & \multicolumn{2}{c|}{\textbf{Windows}}                               & \multicolumn{1}{c|}{\textbf{macOS}}                 & \multicolumn{2}{c|}{\textbf{Manjaro}}                                                          & \multicolumn{1}{c|}{\textbf{Ubuntu}}                & \multicolumn{3}{c|}{\textbf{Android}}                                                                                                                 & \multicolumn{1}{c|}{\textbf{iOS}}              \\ \cline{2-11} 
\multicolumn{1}{|c|}{\multirow{-2}{*}{\textbf{Test / OS}}} & \multicolumn{1}{c|}{\textbf{10}} & \multicolumn{1}{c|}{\textbf{11}} & \multicolumn{1}{c|}{\textbf{Monterey 12.6.7}}       & \multicolumn{1}{c|}{\textbf{KDE 23.0.1}} & \multicolumn{1}{c|}{\textbf{GNOME 23.0.1}}          & \multicolumn{1}{c|}{\textbf{22.04.3 LTS}}           & \multicolumn{1}{c|}{\textbf{8}}                     & \multicolumn{1}{c|}{\textbf{12}}               & \multicolumn{1}{c|}{\textbf{13}}               & \multicolumn{1}{c|}{\textbf{16}}               \\ \hline
\textbf{}                                                  & \textbf{}                        & \textbf{}                        & \textbf{}                                           & \textbf{}                                & \textbf{}                                           & \textbf{}                                           & \textbf{}                                           & \textbf{}                                      & \textbf{}                                      & \textbf{}                                      \\ \hline
\multicolumn{1}{|c|}{\textbf{NET-1}}                       & \multicolumn{1}{c|}{YES}         & \multicolumn{1}{c|}{YES}         & \multicolumn{1}{c|}{YES}                            & \multicolumn{1}{c|}{YES}                 & \multicolumn{1}{c|}{YES}                            & \multicolumn{1}{c|}{YES}                            & \multicolumn{1}{c|}{YES}                            & \multicolumn{1}{c|}{YES}                       & \multicolumn{1}{c|}{YES}                       & \multicolumn{1}{c|}{YES}                       \\ \hline
\multicolumn{1}{|c|}{\textbf{NET-2}}                       & \multicolumn{1}{c|}{YES}         & \multicolumn{1}{c|}{YES}         & \multicolumn{1}{c|}{}                               & \multicolumn{1}{c|}{YES}                 & \multicolumn{1}{c|}{YES}                            & \multicolumn{1}{c|}{YES}                            & \multicolumn{1}{c|}{YES}                            & \multicolumn{1}{c|}{YES}                       & \multicolumn{1}{c|}{YES}                       & \multicolumn{1}{c|}{}                          \\ \hline
\textbf{}                                                  &                                  &                                  &                                                     &                                          &                                                     &                                                     &                                                     &                                                &                                                &                                                \\ \hline
\multicolumn{1}{|c|}{\textbf{WA-1}}                        & \multicolumn{1}{c|}{YES}         & \multicolumn{1}{c|}{YES}         & \multicolumn{1}{c|}{{\color[HTML]{C9211E} NO}}      & \multicolumn{1}{c|}{YES}                 & \multicolumn{1}{c|}{{\color[HTML]{C9211E} NO}}      & \multicolumn{1}{c|}{{\color[HTML]{C9211E} NO}}      & \multicolumn{1}{c|}{{\color[HTML]{C9211E} NO}}      & \multicolumn{1}{c|}{{\color[HTML]{C9211E} NO}} & \multicolumn{1}{c|}{{\color[HTML]{C9211E} NO}} & \multicolumn{1}{c|}{{\color[HTML]{C9211E} NO}} \\ \hline
\multicolumn{1}{|c|}{\textbf{WA-2}}                        & \multicolumn{1}{c|}{YES}         & \multicolumn{1}{c|}{YES}         & \multicolumn{1}{c|}{}                               & \multicolumn{1}{c|}{YES}                 & \multicolumn{1}{c|}{}                               & \multicolumn{1}{c|}{}                               & \multicolumn{1}{c|}{}                               & \multicolumn{1}{c|}{}                          & \multicolumn{1}{c|}{}                          & \multicolumn{1}{c|}{}                          \\ \hline
\multicolumn{1}{|c|}{\textbf{WA-2}}                        & \multicolumn{1}{c|}{YES}         & \multicolumn{1}{c|}{YES}         & \multicolumn{1}{c|}{}                               & \multicolumn{1}{c|}{YES}                 & \multicolumn{1}{c|}{}                               & \multicolumn{1}{c|}{}                               & \multicolumn{1}{c|}{}                               & \multicolumn{1}{c|}{}                          & \multicolumn{1}{c|}{}                          & \multicolumn{1}{c|}{}                          \\ \hline
\textbf{}                                                  &                                  &                                  &                                                     &                                          &                                                     &                                                     &                                                     &                                                &                                                &                                                \\ \hline
\multicolumn{1}{|c|}{\textbf{WB-1}}                        & \multicolumn{1}{c|}{}            & \multicolumn{1}{c|}{}            & \multicolumn{1}{c|}{{\color[HTML]{C9211E} NO $^4$}} & \multicolumn{1}{c|}{}                    & \multicolumn{1}{c|}{{\color[HTML]{C9211E} NO $^3$}} & \multicolumn{1}{c|}{{\color[HTML]{C9211E} NO $^3$}} & \multicolumn{1}{c|}{{\color[HTML]{C9211E} NO $^1$}} & \multicolumn{1}{c|}{YES $^2$}                  & \multicolumn{1}{c|}{YES $^2$}                  & \multicolumn{1}{c|}{YES}                       \\ \hline
\multicolumn{1}{|c|}{\textbf{WB-2}}                        & \multicolumn{1}{c|}{}            & \multicolumn{1}{c|}{}            & \multicolumn{1}{c|}{NotAllowedError}                & \multicolumn{1}{c|}{}                    & \multicolumn{1}{c|}{}                               & \multicolumn{1}{c|}{}                               & \multicolumn{1}{c|}{}                               & \multicolumn{1}{c|}{}                          & \multicolumn{1}{c|}{}                          & \multicolumn{1}{c|}{NotAllowedError}           \\ \hline
\end{tabular}%
}

\vspace{5px}

\begin{tabular}{lll}
$^1$ & Android web browser menu has the option to “open in the web browser”, which allows the user to choose the alternative. &  \\
$^2$ & Before opening the link in an external app, it asks for confirmation to the user with a warning message.               &  \\
$^3$ & The user should manually copy the URL or open another browser manually.                                                &  \\
$^4$ & There is no option for the user to open the URL in another browser manually.                                           & 
\end{tabular}%

\caption{Compatibility of the captive portal solution across different operating systems.}
\label{table:compatibility}
\end{table*}

\subsubsection{Usability experiment with users}
As explained in section~\ref{subsec:validation-usability} we designed and conducted an usability experiment with 15 subjects with their own smartphone devices. Users were asked to connect to the Wi-Fi network running the developed captive portal authentication service, and authenticate with a registered security key.

From the 13 users with compatible smartphones, 69.23\% completed the task of connecting to the Wi-Fi network and successfully authenticate in the captive portal with security keys. Table~\ref{table:results:completeness:step} shows the completeness of every step of the task. As we can see, 84.62\% reached a compatible web browser to start authentication. From those who had a compatible web browser, 81.82\% completed authentication.

\begin{table}[!ht]
\resizebox{\linewidth}{!}{%
\def\arraystretch{1.2}%
\begin{tabular}{|l|r|r|}
\hline
\multicolumn{1}{|c|}{\textbf{Step}} & \multicolumn{1}{c|}{\textbf{Total completeness}} & \multicolumn{1}{c|}{\textbf{Step completeness}} \\ \hline
Connect to Wi-Fi network            & 100.00\%                                         & 100.00\%                                        \\ \hline
Follow OS instructions              & 92.31\%                                          & 92.31\%                                         \\ \hline
Browser redirection                 & 84.62\%                                          & 91.67\%                                         \\ \hline
Authentication                      & 69.23\%                                          & 81.82\%                                         \\ \hline
\end{tabular}%
}
\caption{Completeness of every step of the proposed task. Sample of 13 users with compatible smartphones.}
\label{table:results:completeness:step}
\end{table}

On the other hand, we detected several errors of different categories, that constituted different types of obstacles during the execution of the task. Namely, we found (1) use errors, produced by users; (2) OS errors, caused by the software of the user device; (3) connectivity issues, while connecting the security key via USB or NFC; (4) compatibility errors, as incompatibility with captive portal on iPhone devices.

Table~\ref{table:results:completeness:error-rate} shows the error rate of $1.93$. Most of the detected errors were operating system-related errors. Specifically, the most common error was the Android credential manager menu that delayed to appear, which confused some users. Even further, in some cases we detected that the Android credential manager menu blocks the completion of the operation by getting frozen, causing the user to abandon the task. On the other hand, there were errors caused by the use. We observed that some users tried to connect the key directly to the unlocked smartphone, expecting the Wi-Fi network was automatically connected and authenticated without any additional step. Finally, some of the errors were caused by connectivity of the USB using an adaptor.

\begin{table}[!ht]
\centering
\def\arraystretch{1.2}%
\begin{tabular}{|l|r|}
\hline
\multicolumn{1}{|c|}{\textbf{Type of error}} & \multicolumn{1}{c|}{\textbf{Error rate}} \\ \hline
Use error                                    & 0.67                                     \\ \hline
OS error                                     & 0.73                                     \\ \hline
Connectivity error                           & 0.40                                     \\ \hline
Compatibility error                          & 0.13                                     \\ \hline
Total error rate                             & 1.93                                     \\ \hline
\end{tabular}%
\caption{Error rate of the different types of errors.}
\label{table:results:completeness:error-rate}
\end{table}

\begin{figure}[!ht]
    \centering
    \includegraphics[width=0.9\linewidth]{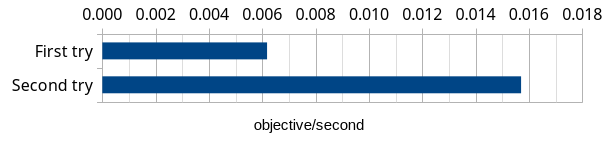}
    \caption{Time-Based Efficiency (TBE). Average objectives per second in each try. Users have learnt from errors in the second try.}
    \label{img:results:graph_time_based_efficiency}
\end{figure}

\begin{figure}[!ht]
    \centering
    \includegraphics[width=0.9\linewidth]{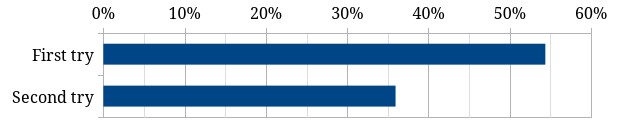}
    \caption{Overall Relative Efficiency (ORE). Ratio of time taken by the users who successfully completed the task in relation to the total time taken by all users.}
    \label{img:results:graph_overall_relative_efficiency}
\end{figure}

Most of the observed errors did not cause the task to fail, although they delayed its completion. For measuring the efficiency, we have calculated Time-Based Efficiency (TBE) and Overall Relative Efficiency (ORE), shown in figures~\ref{img:results:graph_time_based_efficiency} and~\ref{img:results:graph_overall_relative_efficiency}. In the first try, users completed $6.18 \times 10^{-4}$ objectives per second, while in the second try, where the users have learnt from the errors, users completed $1.571 \times 10^{-3}$ objectives per second, more than the double. Also, we measured the Overall Relative Efficiency as the ratio of time taken by the users who successfully completed the task in relation to the total time taken by all users. The ORE of the first try is 54.38\%, and 35,95\% in the second try. In the second try, the users who failed the first time, spent more time than the ones that completed the task in the first time.

\begin{figure}[!ht]
    \centering
    \includegraphics[width=\linewidth]{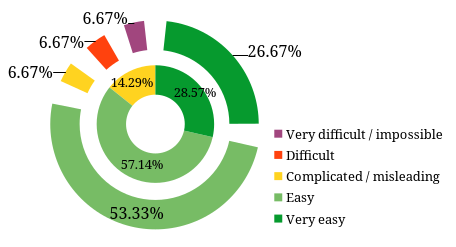}
    \caption{Difficulty as a measure of satisfaction: overall satisfaction (outer) and satisfaction from recovering from errors (inner). Answers from users after performing the task.}
    \label{img:results:graph_satisfaction_overall_vs_error}
\end{figure}

\begin{figure}[!ht]
    \centering
    \includegraphics[width=\linewidth]{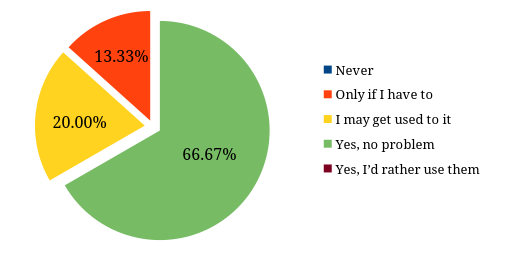}
    \caption{User answers to ``Would you use security keys for connecting to a Wi-Fi network?" after finishing the experiment.}
    \label{img:results:graph_satisfaction_final}
\end{figure}

Finally, we have measured satisfaction through different questions during the experiment. According to figure~\ref{img:results:graph_satisfaction_overall_vs_error}, 53.33\% of users have found easy the process of connecting to the Wi-Fi with security keys, and 57.14\% have found easy to try again when some error occurred. At the end of the experiment, as shown in figure~\ref{img:results:graph_satisfaction_final}, 66.67\% of users have declared that they would have no problem in use security keys for connecting to a Wi-Fi network, but none of them would prefer using keys.

\section{Discussion and conclusions}
\label{sec:discussion}
We propose a novel authentication protocol based on FIDO2 authentication  integrated it with a captive portal. Moreover, we have developed a functional prototype on the OpenNDS captive portal. The prototype can be installed on a network to authenticate users with FIDO2 security keys. In addition, we have validated the prototype usability, finding use errors that can drive future research on FIDO2 usability with a more extensive experiment.

Our FIDO2CAP protocol and the implemented prototype are a proof of how FIDO2 authentication can be integrated with other systems, and serve as a replacement of passwords and their vulnerabilities. Therefore, security keys and passkeys can now be used with captive portals that authenticate users in real network authentication environments. However, there is some work left in relation to the usability and producing mature software with full compatibility across devices and web browsers. Our implemented prototype depends on custom web browsers that operating systems open when detecting a captive portal, impacting the user experience.

In conclusion, FIDO2 authentication represent an opportunity to improve security by migrating the existing authentication services to a new authentication paradigm. Considering that central authentication involves different systems, the migration to FIDO2 should be done in all authentication scenarios of a business. They include web authentication, but also other systems like network access control.

WebAuthn and FIDO CTAP are recent standards that is still under development. Although it has already been implemented in web authentication in different operating systems, browsers and devices, their applicability to further scenarios has not yet been studied. This paper proves that there are other applications, like network authentication. Although the underlining technology is web-based, the integration with a real captive portal enforcement device demonstrates the potential of security keys in network authentication.

\section{Future work}
\label{sec:future-work}

In this paper we proposed FIDO2CAP, and validated it with a real prototype and users. Here we include some of our current research lines that continue the work we presented in this paper:

\begin{enumerate}
    \item \textbf{Validate the usability of the solution with more users and compare different samples}. Design new environments and validate user acceptance across different types of users.
    \item \textbf{Extend the prototype for auto-registration and validate its usability}. Test the usability of users registering their own credentials in the captive portal for later authentication.
    \item \textbf{Compare the usability of the developed captive portal and a captive portal based on passwords}. Evaluate how passwordless authentication with security keys is more usable than passwords.
    \item \textbf{Study the usability issues caused by the error management of web browsers with WebAuthn}. We detected some differences when managing exceptions during a failure in WebAuthn authentication. We are working on a study of the current browser implementations and proposing a solution, which can help to improve usability and adoption of this authentication method.
\end{enumerate}


\section*{Acknowledgements}
This work was supported by the grant ED431C 2022/46 – Competitive Reference Groups GRC – funded by: EU and ``Xunta de Galicia'' (Spain). This work was also supported by CITIC, funded by ``Xunta de Galicia'' through the collaboration agreement between the ``Consellería de Cultura, Educación, Formación Profesional e Universidades'' and the Galician universities to strengthen the research centres of the ``Sistema Universitario de Galicia'' (CIGUS). Also, the work is founded by the ``Formación de Profesorado Universitario'' (FPU) grant from the Spanish Ministry of Universities to Martiño Rivera Dourado (Grant FPU21/04519).

\bibliographystyle{IEEEtran}

\end{document}